\documentclass{jkas}

\def\year{2015} 
\def\month{April} 
\def\volume{48} 
\def\issue{2} 
\def\beginpage{1} 
\def\endpage{9} 
\def\received{February 9} 
\def\accepted{March **, 2015} 
\date{Received \received; accepted \accepted}

\def\ie{{i.e.,\ }}
\def\kms{~{\rm km~s^{-1}}}
\def\cm3{~{\rm cm^{-3}}}

\def\muG{~{\mu\rm G}}

\usepackage{flushend}

\title{Radio Emission from Weak Spherical Shocks in the Outskirts of Galaxy Clusters}

\author{Hyesung Kang}

\affil{Department of Earth Sciences, Pusan National University, Pusan 609-735, Korea; \email{hskang@pusan.ac.kr}}

\def\corrauthor{H. Kang}
\def\runningauthor{Kang}
\def\runningtitle{Radio Emission from Spherical Cluster Shocks}
\def\keywords{acceleration of particles --- cosmic rays ---
galaxies: clusters: general --- shock waves}

\def\abstracttext{
In \citet{kang15} we calculated the acceleration of cosmic-ray electrons and the ensuing radio synchrotron emission  
at weak spherical shocks that are expected to form in the outskirts of galaxy clusters.
There we demonstrated that, at decelerating spherical shocks,
the volume integrated spectra of both electrons and radiation 
deviate significantly from the test-particle power-laws predicted for constant planar shocks, 
because the shock compression ratio and the flux of injected electrons 
decrease in time.
In this study, we consider spherical blast waves propagating into a constant density core 
surrounded by an isothermal halo with $\rho\propto r^{-n}$ in order to explore how the 
deceleration rate of the shock speed affects the radio emission from accelerated electrons.
The surface brightness profile and the volume-integrated radio spectrum of the model shocks are 
calculated by assuming a ribbon-like shock surface on a spherical shell and 
the associated downstream region of relativistic electrons.
If the postshock magnetic field strength is about $7\muG$,
at the shock age of $\sim 50$~Myr, the volume-integrated radio spectrum 
steepens gradually with the spectral index from
$\alpha_{\rm inj}$ to $\alpha_{\rm inj}+0.5$ over 0.1-10~GHz,
where $\alpha_{\rm inj}$ is the injection index at the shock position
expected from the diffusive shock acceleration theory.
Such gradual steepening could explain
the curved radio spectrum of the radio relic in cluster A2266, 
which was interpreted as a broken power-law by \citet{trasatti14}, 
if the relic shock is young enough so that the break frequency falls in around $1$~GHz.
}

\begin{document}
\jkashead 
\section{INTRODUCTION}

Cosmological hydrodynamic simulations have shown that 
shock waves may form due to supersonic flow motions in
the baryonic intracluster medium (ICM) during the formation of the large scale 
structure in the Universe \citep[e.g.,][]{ryu03,kang07,vazza09,skill11}.
The time evolution of the spatial distribution of such shocks in numerical simulations 
\citep[e.g.,][]{vazza12} indicates that shock surfaces behave like spherical bubbles blowing out from the cluster center,
when major episodes of mergers or infalls from adjacent filaments occur \.
Shock surfaces seem to last only for a fraction of a dynamical time scale of clusters, 
i.e., $\sim 0.1 t_{\rm dyn}\sim 100$~Myr.
This implies that some cosmological shock waves are associated with merger induced outflows, 
and that spherical shock bubbles are likely to expand into the cluster outskirts with a decreasing density profile.

Observational evidence for shock bubbles can be found at
the so-called ``radio relics'' detected in the outskirts of galaxy cluster,
which are interpreted as synchrotron emitting structures containing relativistic electrons
accelerated at weak ICM shocks ($M_s \sim 2-4$) \citep[e.g.,][]{vanweeren10, vanweeren11, nuza12, feretti12,brunetti2014}. 
Some radio relics, for instance, the ``Sausage relic'' in galaxy cluster CIZA J2242.8+5301 and the
``Toothbrush'' relic in galaxy cluster 1RXS J0603.3+4214, have thin arc-like shapes of $\sim 50$~kpc 
in width and $\sim 1-2$ Mpc in length \citep{vanweeren10, vanweeren12}. 
They could be represented by a ribbon-like structure on a spherical shell
projected onto the sky plane \citep[e.g.,][]{vanweeren10,kang12}.

As shown in Fig. 1, 
the viewing depth of a relic shock structure can be parameterized by the extension angle $2\psi$,
while the physical width of the postshock shell of radiating electrons 
is mainly determined by the advection length, 
$\Delta l_{\rm adv}(\gamma_e) \approx u_2 \cdot \rm{min}[t_{\rm age},t_{\rm rad}(\gamma_e)]$,
where $u_2$ is the flow speed behind the shock and $t_{\rm rad}(\gamma_e)$ is the radiative
cooling time scale for electrons with Lorentz factor, $\gamma_e$.
It remains largely unknown how such a ribbon-like structure can be formed by
the spherical outflows in galaxy clusters.
We note, however, a recent study by \citet{shimwell15} who suggested that the {\it uniform} arc-like
shape of some radio relics may trace the underlying region of pre-existing seed electrons 
remaining from an old radio lobe.

\begin{figure*}[t]
\centering
\vskip 0.0cm
\includegraphics[width=120mm]{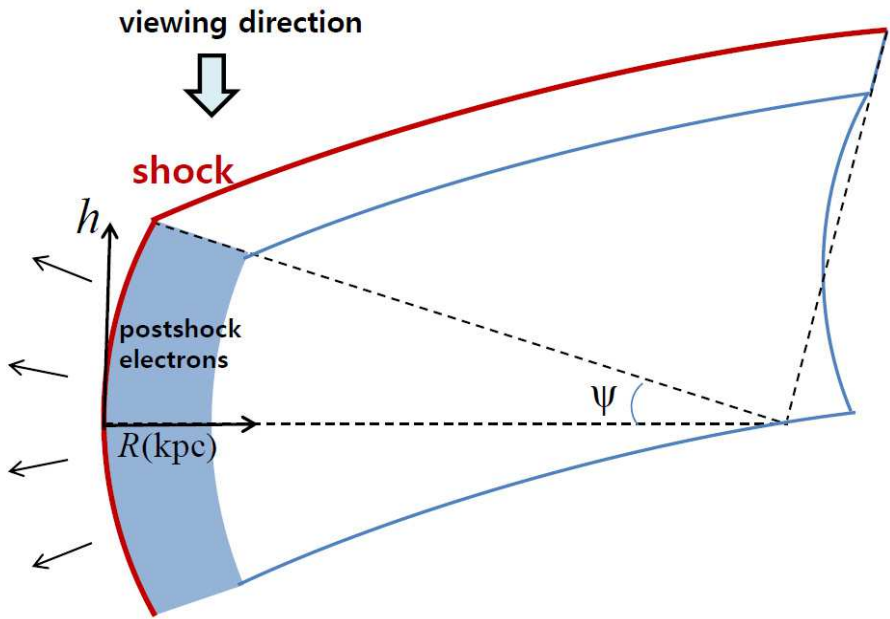}
\vskip 0.0cm
\caption{ 
Schematic diagram showing a spherical shock surface (red) and the postshock electron distribution behind the shock adopted in this study.
The shock surface is modeled as a Mpc-long ribbon on the spherical surface whose viewing depth can be defined by the 
extension angle $2\psi$.
The depth of the postshock electron distribution behind the shock is determined by the advection length, $\Delta l_{\rm adv}\approx u_2 \cdot t_{\rm age}$, for low energy electrons
or the cooling length, $\Delta l_{\rm cool}(\gamma_e)\approx u_2 \cdot t_{rad}(\gamma_e)$
for high energy electrons.
The radio emissivity from CR electrons, $j_{\nu}(d)$ is calculated using the DSA simulation results in Paper I (where $d$ is the distance from the shock).
The surface brightness is calculated by integrated $j_{\nu}(d)$ over the path along 
the line of sight (see Eq. [\ref{IR}]).
If viewed from the top as indicated here, the projected image will have an arc-like shape similar to the 
Sausage relic in CIZA J2242.8+5301 \citep{vanweeren10}.
}
\end{figure*}

According to diffusive shock acceleration (DSA) theory, 
cosmic-ray (CR) particles can be generated 
via Fermi 1st order process  at collisionless shocks \citep{bell78, dru83, maldru01}.
The {\it test-particle} DSA theory predicts that the CR energy spectrum at the shock
position has a power-law energy spectrum, $N(E) \propto E^{-s}$,
where $s=(\sigma+2)/(\sigma-1)$ and $\sigma=\rho_2/\rho_1$ is the shock compression ratio.
Hereafter, we use the subscripts `1' and `2' to denote the
conditions upstream and downstream of shock, respectively.
Then the synchrotron radiation spectrum due to CR electrons at the shock location has a power-law form of
$j_{\nu}(x_s)\propto \nu^{-\alpha_{\rm inj}}$, where $\alpha_{\rm inj} = (s-1)/2$ is
the injection index.
Moreover, the volume-integrated synchrotron spectrum downstream
of a planar shock becomes a simple power-law of $J_{\nu} \propto \nu^{-A_{\nu}}$
with $A_{\nu}=\alpha_{\rm inj}+0.5$ above a break frequency, 
since electrons cool via synchrotron and inverse-Compton (iC) losses behind the shock \citep[e.g.,][]{kang11}.
Such spectral characteristics are commonly used to infer the shock properties 
(i.e., Mach number) of observed radio relic shocks \citep[e.g.,][]{vanweeren10,stroe14}

In \citet{kang15} (Paper I), we calculated the electron acceleration at spherical shocks 
similar to Sedov-Taylor blast waves with $M_s\sim 2.5-4.5$, which expand into a hot {\it uniform} ICM.
We found that the electron energy spectrum at the shock location has reached 
the steady state defined by the {\it instantaneous} shock parameters. 
So the spatially resolved, synchrotron radiation spectra at the shock 
could be described properly 
by the test-particle DSA predictions for planar shocks.
However, the volume integrated spectra of both electrons and radiation evolve differently 
from those of planar shocks and exhibit some nonlinear signatures, 
depending on the time-dependent evolution of the shock parameters.
For instance, the shock compression ratio $\sigma$
and the injection flux of CR electrons decrease in time,
as the spherical shock expands and slows down,
resulting in some curvatures in both electron and radiation spectra.

Magnetic fields play key roles in DSA at collisionless shocks 
and control the synchrotron cooling and emission of relativistic electrons.
Observed magnetic field strength is found to decrease from $\sim 1-10 \muG$ in the core region to
$\sim 0.1-1 \muG$ in the periphery of clusters \citep{feretti12}.
On the other hand, it is well established that magnetic fields can be amplified via resonant 
and non-resonant instabilities
induced by CR protons streaming upstream of strong shocks \citep{bell78,lucek00,bell04}.
Recently, \citet{capri14} has shown that the magnetic field amplification (MFA)
factor scales with the Alfv\'enic Mach number, $M_A$, and the CR proton acceleration efficiency 
as $\langle \delta B/B\rangle^2 \sim$ $3 M_A (P_{\rm cr,2}/\rho_1 u_s^2)$.
Here $\delta B$ is the turbulent magnetic fields perpendicular to the mean background magnetic fields,
$\rho_1$ is the upstream gas density,
and $P_{\rm cr,2}$ is the downstream CR pressure.
For typical cluster shocks with $3\lesssim M_s \lesssim 5$ and $M_A\sim 10$ \citep{ryu03}, 
the MFA factor due to the streaming stabilities is expected be rather small but not negligible, 
$\langle \delta B/B\rangle^2 \sim 0.3-1.5$.
However, it has not yet been fully understood how magnetic fields may be amplified in both upstream 
and downstream of a weak shock in high beta ICM plasmas with $\beta_p=P_g/P_B \sim 100$.

Therefore, in Paper I, we considered several models with decaying postshock magnetic fields, 
in which the downstream magnetic field, $B_d(r)$, decreases behind the shock with a scale height of 100-150~kpc.
We found that the impacts of different $B_d(r)$ profiles 
on the spatial distribution of the electron energy spectrum, $N_e(r,\gamma_e)$,
is not substantial, because iC scattering off cosmic background photons 
provides the baseline cooling rate.
Any variations in $N_e(r,\gamma_e)$ is smoothed out in the spatial distribution of 
the synchrotron emissivity, $j_{\nu}(r)$, because electrons in a broad range of 
$\gamma_e$ contribute to $j_{\nu}$ at a given frequency.
We note, however, that the $B^2$ dependence of the synchrotron emissivity
($j_{\nu}\propto N_eB^2$) can become significant in some cases.
Moreover, any nonlinear features due to the spatial variations of $N_e(r,\gamma_e)$
and $B_d(r)$ are mostly averaged out, 
leaving only subtle signatures in the volume integrated spectrum, $J_{\nu}$.

For the case with a constant background density (e.g., {\bf MF1-}3 model in Paper 1), the shock speed decreases
approximately as $u_s\propto t^{-3/5}$.
In fact, it has not been examined, through cosmological hydrodynamic simulations,
whether these shock bubbles would accelerate or decelerate
as they propagate into the cluster outskirts with a decreasing density profile.
In this study, we have performed additional DSA simulations, in which 
the initial Sedov blast wave propagates into the background medium with 
$\rho_u \propto r^{-n}$, where $r$ is the radial distance from the cluster center
and $n=2-4$.
This effectively mimics a blast wave that expands into a constant-density core surrounded 
by an isothermal halo with a decreasing pressure profile.
In these new runs, the spherical shock decelerates much slowly, compared to $u_s\propto t^{-3/5}$,
and the nonlinear effects due to the deceleration of the shock speed are expected to be
reduced from what we observed in the uniform density models considered in Paper I.

Here we also have calculated the projected surface brightness, $I_{\nu}$,
which depend on the three dimensional structure of the shock surfaces and the viewing direction.
Because of the curvature in the model shock structure,
synchrotron emissions from downstream electrons with different ages contribute to the surface 
brightness along a given line-of-sight (\ie projection effects).
So the observed spatial profile of $I_{\nu}(R)$ is calculated by assuming
the geometrical configuration described in Fig. 1. 
Note that this model with a ribbon-like shock surface gave rise to radio flux profiles, $S_{\nu}(R)$,
that were consistent with those of the the Sausage relic in CIZA J2242.8+5301 and the double relics in 
ZwCl0008.8+5215 \citep[e.g.,][]{vanweeren10, vanweeren11, kang12}.

In paper I, we demonstrated that the spectral index of the volume-integrated spectral index 
increases gradually from $A_{\nu}=\alpha_{\rm inj}$ to $A_{\nu}=\alpha_{\rm inj}+0.5$
over a broad frequency range, $\sim (0.1-10)\nu_{\rm br}$,
where the break frequency is $\nu_{\rm br}\sim 0.5$~GHz at the shock age of about 50~Myr
for the postshock magnetic fields $\sim7\muG$ (see Eq. [\ref{fbr}]). 
Here we have explored whether such a transition can explain the broken power-law spectra 
observed in the radio relic in A2256 \citep{trasatti14}.
 
In the next section we describe the numerical calculations. 
The DSA simulation results of blast wave models with different postshock magnetic 
field profiles and with different background density profiles will be discussed in 
Section 3.
A brief summary will be given in Section 4.

\section{Numerical Calculations}

In order to calculate DSA of CR electrons at spherical shocks,
we have solved the time-dependent diffusion-convection equation
for the pitch-angle-averaged phase space distribution function
for CR electrons, $f_e(r,p,t)$, in the one-dimensional spherically symmetric geometry \citep{skill75}.
The test-particle version of CRASH (Cosmic-Ray Amr SHock) code in a comoving spherical grid
was used \citep{kj06}.
The details of the simulation set-up can be found in Paper 1.

For the initial conditions, we adopt a Sedov-Taylor similarity solution propagating into 
a uniform ICM with the following parameters:
the ICM density, $n_{H,1}=10^{-3}\cm3$, the ICM temperature, $T_1=5\times 10^7$K,
the initial shock radius, $r_{s,i}=0.78{\rm Mpc}$, and
the initial shock speed, $u_{s,i}=4.5\times 10^3\kms$
with the sonic Mach number, $M_{s,i}=4.3$ at the onset of the simulation.
The shock parameters change in time as the spherical shock expands out, 
depending on the upstream conditions in the cluster outskirts.
For iC cooling, a redshift $z=0.2$ is chosen as a reference epoch,
so $B_{\rm rad}=3.24\muG \cdot (1+z)^2=4.7\muG$.

The synchrotron emissivity, $j_{\nu}(r)$, at each shell 
is calculated (in units of ${\rm erg~cm^{-3}~s^{-1}~Hz^{-1}~str^{-1}}$), 
using the electron distribution function, $f_e(r,p,t)$, and the magnetic
field profile, $B(r,t)$, from the DSA simulations. 
Then the radio intensity or surface brightness, 
$I_{\nu}\ ({\rm erg~cm^{-2}~s^{-1}~Hz^{-1}~str^{-1}})$, is calculated by integrating $j_{\nu}$ 
along the path length, $h$, as shown in Fig.~1:
\begin{equation}
I_{\nu}(R)= 2 \int_0^{h_{\rm max}} j_{\nu}(d) d {\it h}. 
\label{IR}
\end{equation}
Here $R$ is the distance behind the projected shock edge in the plane of the sky
and $d=r_s-r$ is the distance of a shell behind the shock, where $(r_s-d)^2 = (r_s-R)^2 + h^2$.
The extension angle $\psi=10^{\circ}$ is assumed in this study.

The volume integrated emissivity, $J_{\nu}=\int j_{\nu}(r) dV$, 
is calculated by integrating $j_{\nu}$
over the downstream volume defined in Fig.~1.

The spectral indexes of $j_{\nu}(d)$, $J_{\nu}$, and $I_{\nu}(R)$ are defined as follows:
\begin{eqnarray}
\alpha_{\nu_i-\nu_{i+1}}(d) =- {{d\ln j_{\nu}(d) }\over {d\ln \nu}},  \\ 
A_{\nu_i-\nu_{i+1}} =- {{d\ln J_{\nu} }\over {d\ln \nu}}, \\
B_{\nu_i-\nu_{i+1}} (R) =- {{d\ln I_{\nu}(R) }\over {d\ln \nu}}, 
\end{eqnarray}  
estimated between $\nu_i$ and $\nu_{i+1}$. 
We chose the following three frequencies at the source,
$\nu_1 = 240$~MHz, $\nu_2=600$~MHz, $\nu_3=1.4$~GHz in Figs. 2 and 4.
Then the redshifted frequency for objects at a redshift $z$ is 
$\nu_{\rm obs,i}=\nu_i/(1+z)$.

\begin{table}
\begin{center}
\vskip 0.2cm
{\bf Table 1.}~~Model Parameters\\
\vskip 0.3cm
\begin{tabular}{ lrr }
\hline\hline

Model & $\rho_u(r)$ & $B_d(r)$ \\
\hline

{\bf MF1}$^{\rm a}$  & $\rho_u=\rho_0$ & $B_d=7\mu$G  \\
{\bf MF2}  & $\rho_u=\rho_0$ & $B_d\propto \rho(r)$ \\
\smallskip
{\bf MF3}  & $\rho_u=\rho_0$ & $B_d\propto \sqrt {P(r)}$ \\

{\bf BD1}$^{\rm a}$  & $\rho_u=\rho_0$ & $B_d=7\mu$G  \\
{\bf BD2}  & $\rho_u\propto r^{-2}$ & $B_d=7\mu$G  \\ 
{\bf BD2b} & $\rho_u\propto r^{-2}$ & $B_d\propto \sqrt {P(r)}$  \\
{\bf BD3}  & $\rho_u\propto r^{-4}$ & $B_d=7\mu$G   \\
{\bf BD3b}  & $\rho_u\propto r^{-4}$ & $B_d\propto \sqrt {P(r)}$   \\

\hline
\end{tabular}
\end{center}
$^{\rm a}$  In fact {\bf MF1} and {\bf BD1} models are identical.\\
\end{table}

\section{DSA SIMULATION RESULTS}

\begin{figure*}[t]
\centering
\vskip -0.5cm
\includegraphics[width=140mm]{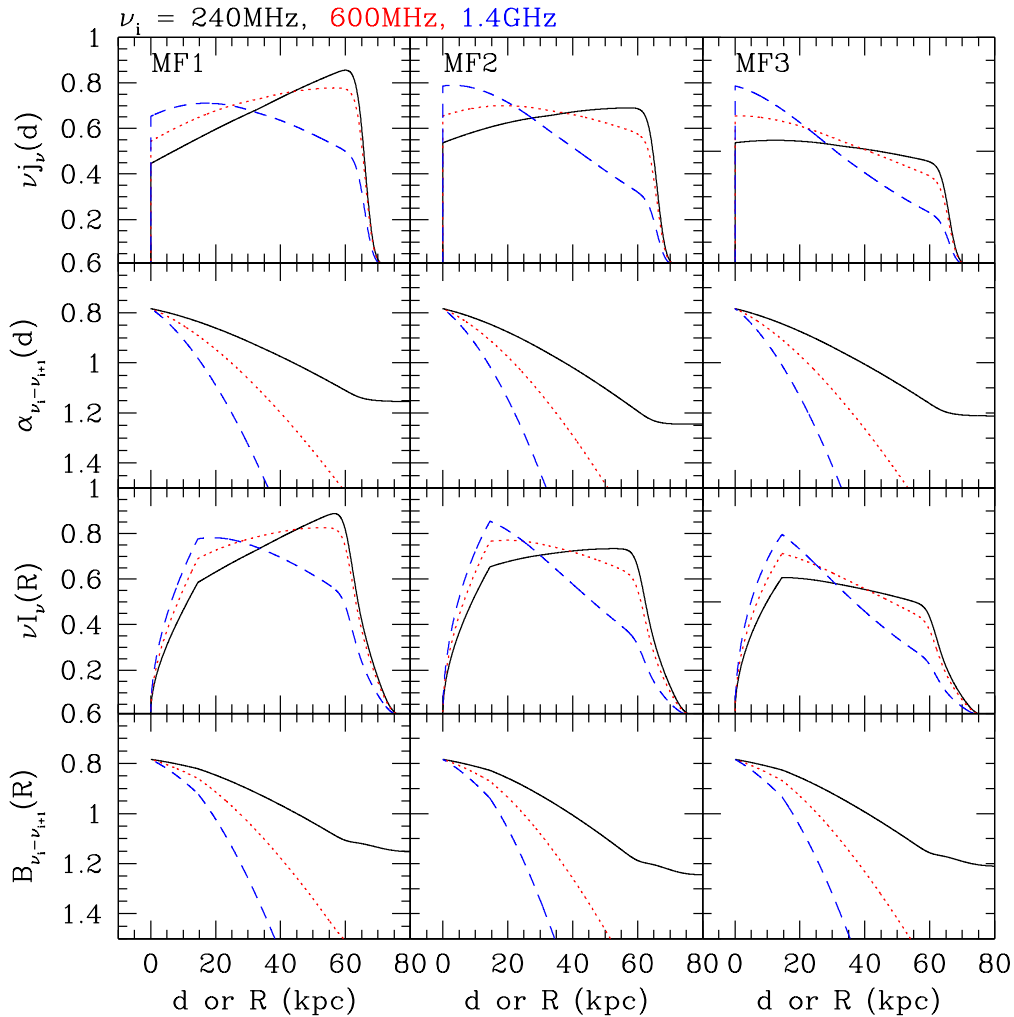}
\vskip -0.0cm
\caption{ 
Spherical shock models with three different magnetic field profiles: {\bf MF1}, {\bf MF2}, and {\bf MF3}. The results are shown at
$t_{\rm age}=47$~Myr, when $r_s=0.96$Mpc, $u_s=3.4\times 10^3 \kms$ and $M_s=3.2$.
In the upper two rows, spatial distributions of the synchrotron emissivity, $\nu_i j_{\nu_i}(r)$, 
and its spectral indexes, $\alpha_{\nu_i-\nu_{i+1}(r)}$ are plotted as a function of the postshock distance $d$ 
from the shock surface.
In the lower two rows, the intensity $\nu_i I_{\nu_i}(R)$
and its spectral indexes, $B_{\nu_i-\nu_{i+1}}(R)$ are plotted as
a function of the {\it projected} distance $R$ from the shock.
The frequency is $\nu_i=$ 240~MHz (black solid line), 600~MHz (red dotted), 
1.4~GHz (blue dashed), and 3~GHz for $i=1$, 2, 3, and 4.
The downstream volume of radio-emitting electrons is assumed to have the same shape as
the one illustrated in Fig. 1 with the extension angle $\psi=10^{\circ}$.
Note that $j_{\nu}$ and $I_{\nu}$ are plotted in arbitrary units.
}
\end{figure*}
\subsection{Shocks with Different Magnetic Field Profiles}

As in the previous study of \citet{kang12}, we adopt the postshock magnetic field strength
$B_2\sim 7\muG$ in order to explain the observed width of the Sausage relic ($\Delta l \sim 50$~kpc),
while the preshock magnetic field strength is chosen to be $B_1\sim 2-3 \muG$.
Since $B_1$ cannot be constrained directly from observations,
it is adjusted so that $B_2$ becomes about 7 $\muG$ after considering MFA or
compression of the perpendicular components of magnetic fields across the shock.

\begin{figure*}[t]
\centering
\vskip -0.5cm
\includegraphics[width=140mm]{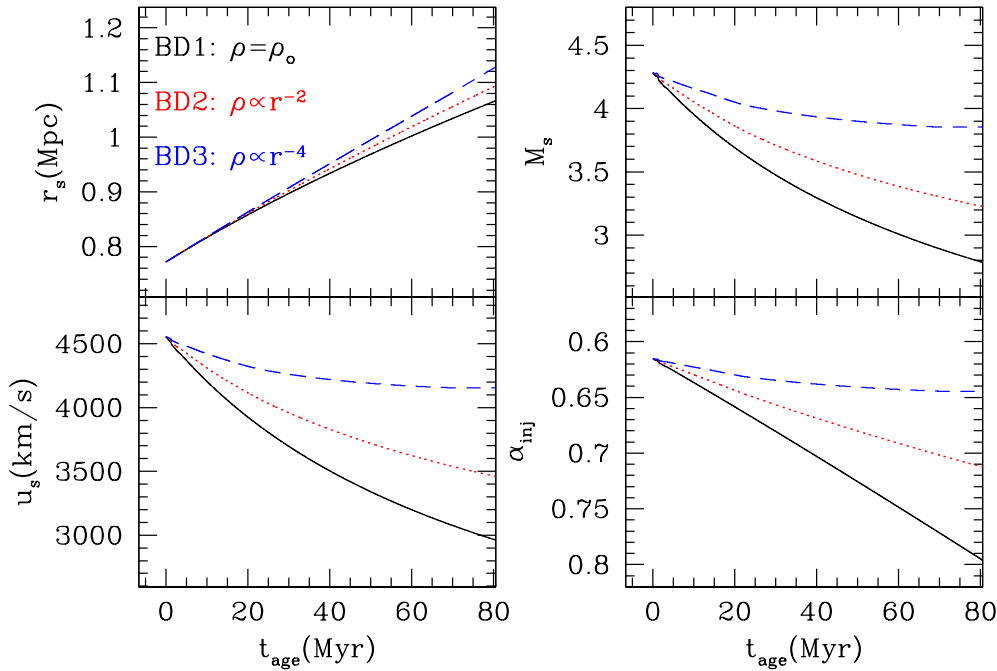}
\vskip -4.0cm
\caption{ 
Evolution of spherical shock models with different background density profile, {\bf BD1} (black solid lines), 
{\bf BD2} (red dotted), and {\bf BD3} (blue dashed): shock radius, $R_s$, shock speed, $u_s$, 
sonic Mach number, $M_s$, and the DSA spectral index at the shock, $\alpha_{\rm inj}$. 
}
\end{figure*}

We consider several models whose characteristics are summarized in Table 1.
As described in Paper I,
we adopt the following downstream magnetic field profile, $B_d(r)$ in {\bf MF} models,
in which the upstream density and temperature are uniform:

{\bf MF1}: $B_1=2\muG$ \& $B_2=7\muG$.

{\bf MF2}: $B_1=3\muG$, $B_2=B_1 \sqrt{1/3+2\sigma^2/3} $,

~~~~~~~~\& $B_d(r)= B_2 \cdot (P(r)/P_2)^{1/2}$ for $r<r_s$.

{\bf MF3}: $B_1=3\muG$, $B_2=B_1 \sqrt{1/3+2\sigma^2/3}$, 

~~~~~~~~\& $B_d(r)= B_2 \cdot (\rho(r)/\rho_2)$ for $r<r_s$.

In Fig. 2 we compare these models at the shock age of 47 Myr. 
Note the evolution of the shock
(\ie $r_s(t)\propto t^{2/5}$ and $u_s(t)\propto t^{-3/5}$) is identical in the three models.
Synchrotron emissivity scales with the electron energy spectrum and the magnetic field strength as $j_{\nu} \propto N_e(\gamma_e)B^2$. 
Of course, the evolution of $N_e(\gamma_e)$ in each {\bf MF} model also depends on 
the assumed profile of $B_d(r)$ through DSA and synchrotron cooling.
In {\bf MF1} model where $B_d(r)=B_2$, the spatial distribution of $j_{\nu}(d)$ at 240~MHz (black solid line)
reveals dramatically the {\it shock  deceleration effects}.
It increases downstream behind the shock because of higher shock compression and higher electron injection flux at earlier time. 
On the other hand, $j_{\nu}(d)$ at 1.4~GHz (blue dashed line) decreases downstream
behind the shock due to the fast synchrotron/iC cooling of high energy electrons.

The downstream increase of $j_{\nu}(d)$ at low frequencies is softened in the models with decaying
postshock magnetic fields because of $B^2$ dependence of the synchrotron emissivity.
In {\bf MF3} model, in which the downstream magnetic field decreases with the gas density behind the shocks, the spatial distributions of
$j_{\nu}(d)$ at all three frequencies decrease downstream.
Thus the downstream distribution of $j_{\nu}(d)$ depends on the shock speed evolution, 
$u_s(t)$ and the postshock $B_d(r)$ as well as the chosen frequency.
In Paper I, we indicated, rather hastily, that signatures imprinted on 
synchrotron emission, $j_{\nu}(r)$, and its volume integrated spectrum, $J_{\nu}$,
due to different postshock magnetic field profiles could be too subtle to detect.
The main reason for this discrepancy is that we examined $\log j_{\nu}(r)$ in Paper I
(see Fig. 6 there), while $\nu j_{\nu}(d)$ is plotted in Fig. 2.

As can be seen in the lower two rows of panels in Fig. 2,
the surface brightness profiles, $I_{\nu}(R)$, are affected by the projection effects
as well as the evolution of $u_s(t)$ and the spatial variation of $B_d(r)$.
For instance, the gradual increase of $I_{\nu}$ just behind the shock up to $R_{\rm inf}<15$~kpc is due to the 
increase of the path length, and its inflection point, $R_{\rm inf}=r_s(1- \cos \psi)$,
depends on the value of $\psi$.
Beyond the inflection point, 
the path length decreases but $j_{\nu}(d)$ may increase or decrease depending on $u_s(t)$ and $B_d(r)$,
resulting in a wide range of spatial profiles of $I_{\nu}(R)$.

\begin{figure*}[t]
\centering
\vskip -0.5cm
\includegraphics[width=140mm]{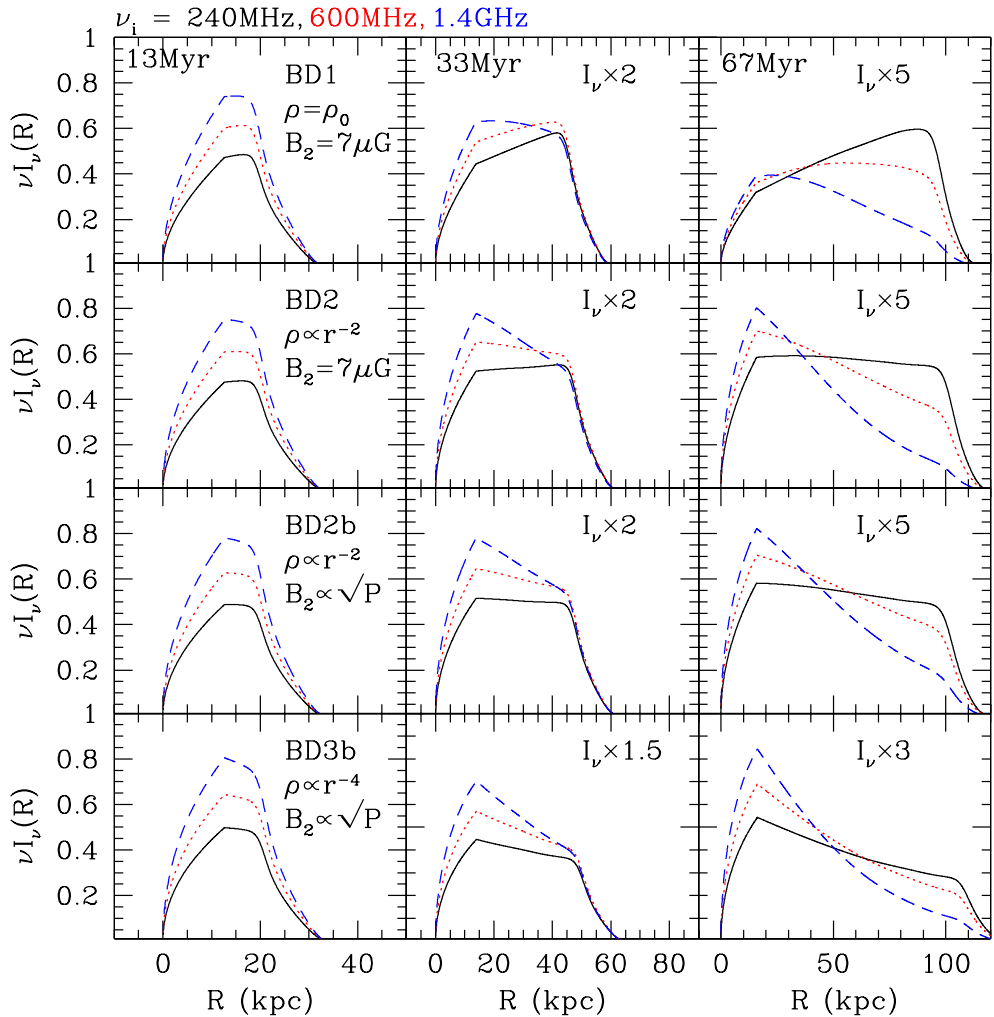}
\vskip -0.2cm
\caption{ 
Spherical shock models with different background density and magnetic field profiles: {\bf BD1} ({\bf MF1}), {\bf BD2}, {\bf BD2b}, and {\bf BD3b} (from top to bottom panels). See Table 1 for the model parameters.
Spatial distributions of the intensity $\nu I_{\nu}$ are shown at
three shock ages, $t_1=13$~Myr, $t_2=33$~Myr, and $t_3=67$~Myr.
The frequency is $\nu_1=240$~MHz (black solid line), $\nu_2=600$~MHz (red dotted), and 
$\nu_1=1.4$~GHz (blue dashed).
The downstream volume of radio-emitting electrons is assumed to have the same shape as
the one illustrated in Fig. 1 with the extension angle $\psi=10^{\circ}$.
The intensity, $\nu I_{\nu}\cdot X$, is plotted in an arbitrary unit, where the numerical factor, $1\le X\le 5$, is
adopted in order to plot the quantities in the linear scale.
}
\end{figure*}

Fig. 2 also demonstrate that
the spectral indexes, $\alpha_{\nu_i-\nu_{i+1}}(d)$ and $B_{\nu_i-\nu_{i+1}}(R)$,
at all three frequencies decrease behind the shock
and do not show significant variations among the different {\bf MF} models
other than faster steepening of both indexes for a more rapid decay of $B_d(r)$. 

In summary, at Sedov-Taylor type spherical shocks decelerating with $u_s\propto t^{-3/5}$,
the distribution function, $N_e(r,\gamma_e)$, of low energy electrons increases downstream behind the shock.
As a result,
the spatial distributions of the radio emissivity, $j_{\nu}(d)$, and
the surface brightness, $I_{\nu}(R)$, at low radio frequencies ($<1$~GHz)
could depend significantly on the postshock magnetic field profile. 
At high radio frequencies, such dependence becomes relatively weaker,
because the postshock width of high energy electrons is much narrower and so
the magnetic field profile far downstream has less influence on synchrotron emission.
On the other hand, the spectral indexes, $\alpha_{\nu}$ and $B_{\nu}$, are relatively insensitive 
to those variations.

\subsection{Shocks with Different Background Density Profile}

\begin{figure*}[t]
\centering
\vskip -0.5cm
\includegraphics[width=140mm]{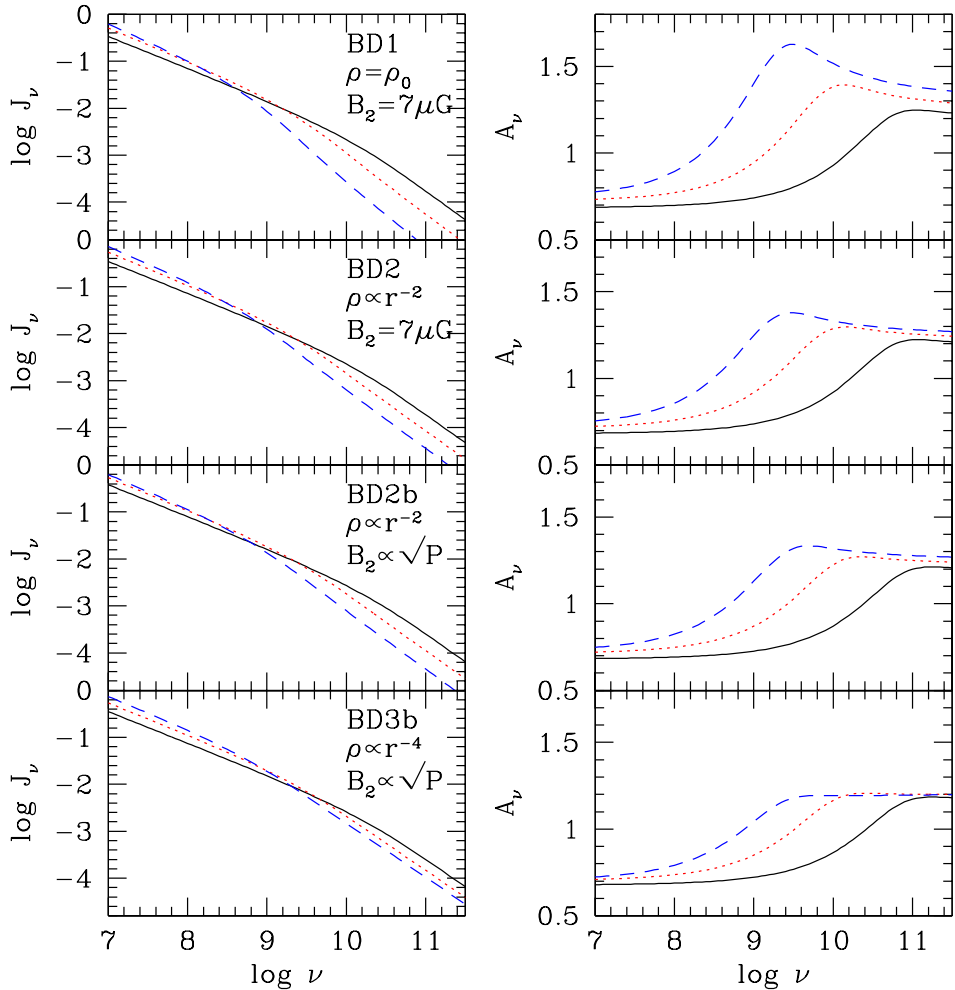}
\vskip -0.2cm
\caption{ 
Same models shown as in Fig. 4.
Volume integrated emissivity, $J_{\nu}$, and its spectral index, $A_{\nu}$, are shown at
three different shock age,
$t_1=13$~Myr (black solid lines),
$t_2=33$~Myr (red dotted lines),
and $t_3=67$~Myr (blue dashed lines).
}
\end{figure*}

In {\bf BD} models, we assume that the initial blast wave propagates into 
an isothermal halo with a different density profile ($r>r_{s,i}$): 

{\bf BD1}: $\rho_u(r)=\rho_0$. 

{\bf BD2}: $\rho_u(r)=\rho_0 \cdot ({r/ r_{s,i}})^{-2}$.

{\bf BD3}: $\rho_u(r)=\rho_0 \cdot ({r/ r_{s,i}})^{-4}$.

Again the upstream temperature ($T_1=5\times 10^7$~K) is uniform, and $B_1=2\muG$ \& $B_2=7\muG$
for these models.
In fact {\bf MF1} and {\bf BD1} models are identical.
We also consider {\bf BD2b} and {\bf BD3b} models, in which the downstream
magnetic field profile is the same as {\bf MF2}, i.e., $B_d(r)\propto \sqrt{P(r)}$.

In the so-call beta model for the gas distribution for isothermal ICM,
$\rho(r)\propto r^{-3\beta}$ in the outskirts of galaxy clusters \citep{sarazin88}.
So {\bf BD2} model corresponds to the beta model 
with $\beta\approx 2/3$, which is consistent with typical X-ray brightness profile 
of observed X-ray clusters.
Recall that in our simulations the spherical blast wave into a uniform ICM is adopted for the initial conditions.
So we are effectively considering a spherical blast wave that propagates into a constant-density core 
(i.e., $\rho=\rho_0$ for $r<r_{s,i}$)
surrounded by a isothermal halo with $\rho\propto r^{-n}$ (for $r>r_{s,i}$).

With the different background density (or gas pressure) profile,
the shock speed evolves differently as it expands outward.
Fig. 3 shows how the shock radius, $r_s(t)$, shock speed,
$u_s(t)$, the sonic Mach number, $M_s(t)$, and the DSA spectral index at the shock position,
$\alpha_{\rm inj}= 3/[2(\sigma-1)]$, varies in time in {\bf BD} models.
As expected, the shock decelerates much slowly if the background pressure declines outward
as in {\bf BD2} and {\bf BD3} models.
In fact, the shock speed is almost constant in {\bf BD3} model.
Hence these {\bf BD} models allow us to explore the dependence of radio spectral properties 
for a wide range of the time evolution of the shock speed.

Fig. 4 compares the surface brightness profile, $I_{\nu}(R)$, at $\nu_1=240$~MHz (black solid lines),
$\nu_2=600$~MHz (red dotted), and $\nu_3=1.4$~GHz (blue dashed) in the models
with different $\rho_u$ and $B_d$.
In {\bf BD1} model, the shock parameters are: 
$R_s=0.83$Mpc, $u_s=4.1\times 10^3 \kms$ and $M_s=3.9$ at $t_1=13$~Myr;
$R_s=0.91$Mpc, $u_s=3.6\times 10^3 \kms$ and $M_s=3.4$ at $t_2=33$~Myr;
$R_s=1.0$Mpc, $u_s=3.1\times 10^3 \kms$ and $M_s=2.9$ at  $t_3=67$~Myr.
Again, we can see that the surface brightness profile of $I_{\nu}(R)$ depends on 
the evolution of $u_s(t)$ and the spatial profile of $B_d(r)$.

The width of the radio structure, $\Delta l_{\rm adv}\sim t_{\rm age} u_s/\sigma$, 
increases with the shock age,
while the amplitude of the surface brightness decreases in time (from left to right in Fig. 4).
For high frequencies ($ > 1$~GHz), however, the width asymptotes to the cooling length,
$\Delta l_{\rm cool}\sim t_{\rm rad}(\gamma_e) u_s/\sigma$.
So the postshock magnetic field strength can be obtained through $t_{\rm rad}(\gamma_e)$
from high frequency observations.
At low frequencies ($<1$~GHz) the surface brightness is affected 
by the evolution of $u_s(t)$ and the spatial variation of $B_d(r)$
as well as the projection effects,
resulting in a wide range of spatial profiles of $I_{\nu}(R)$.
This suggests that, if the radio surface brightness can be spatially resolved 
at several radio frequencies over $\sim (0.1-10)$~GHz, 
we may extract the time evolution of $u_s(t)$ as well as the shock age
from low frequency observations.   
So it would be useful to compare multi-frequency radio observations with 
the intensity modeling that accounts for both shock evolution and projection effects.

The comparison of $I_{\nu}$ of {\bf MF1} ({\bf BD1}) and {\bf MF2} models (see Fig. 2) demonstrates that
the decaying postshock magnetic field could lessen the shock deceleration signatures in the 
surface brightness profile.  
If we compare $I_{\nu}$ of {\bf BD2} and {\bf BD2b} models in Fig. 4, on the other hand, 
the difference in their profiles is much smaller than that between {\bf MF1} and {\bf MF2} models.
This is because the the shock deceleration effects and the ensuing downstream increase 
of $N_e(r,\gamma_e)$ are relatively milder in {\bf BD2} and {\bf BD2b} models.

Next Fig. 5 compares the volume integrated spectrum, $J_{\nu}$, and its spectral index, $A_{\nu}$,
at three different shock ages, $t_{\rm age}=$ 13~Myr (black solid), 33~Myr~(red dotted), and 67~Myr (blue dashed) 
for the same set of models shown in Fig. 4. 

For the test-particle power-law at planar shocks,
the radio index $A_{\nu}$ is the same as $\alpha_{\rm inj}$ for $\nu< \nu_{\rm br}$, where
\begin{equation}
\nu_{\rm br}\approx 0.44 {\rm GHz} \left( {t_{\rm age} \over {50 \rm Myr}} \right)^{-2}
 \left( {B_{e,2} \over {8.4 \muG}} \right)^{-4} \left( {B_{2} \over {7 \muG}} \right)
\label{fbr}
\end{equation}
at the source. 
For $\nu> \nu_{\rm br}$,  $A_{\nu}$ is expected to steepen to $\alpha_{\rm inj}+ 0.5$ 
due to synchrotron/iC cooling of electrons.
Note that the smallest possible index is $\alpha_{\rm inj}=0.5$ at strong shocks, so 
$A_{\nu}=\alpha_{\rm inj}+0.5 \ge 1.0$ above the break frequency.
Also note that the observed (redshifted) break frequency corresponds 
to $\nu_{\rm br,obs}= \nu_{\rm br}/(1+z)$.
If the shock age is $t_{\rm age}\sim 67$~Myr and $B_2\sim 7 \muG$, 
the observed break frequency becomes about 120~MHz for objects at $z=0.2$. 

As can be seen in Fig. 5, however, the transition from $\alpha_{\rm inj}$ to $\alpha_{\rm inj}+0.5$
occurs gradually over about two orders of magnitude in frequency range.
If the shock is 30-70 Myr old for the shock parameters considered here, 
the volume-integrated radio spectrum is expected to
steepen gradually from 100~MHz to 10~GHz, instead of a sharp broken power-law 
with the break frequency at $\nu_{\rm br,obs}$. 
Such gradual steepening may explain why the volume-integrated radio spectrum of 
some relics 
was interpreted as a broken power-law with $A_{\nu}<1.0$ at low frequencies.

According to a recent observation of the relic in A2256, for example, 
the observed spectral index is $A_{351}^{1369} \approx 0.85$ 
between 351 and 1369~MHz and increases to $A_{1369}^{10450} \approx 1.0$ between 1369 and 10450~MHz
\citep{trasatti14}.
We estimated similar indexes for our $J_{\nu}$ between 355 and 1413~MHz, 
and between 1413~MHz and 10~GHz, using the DSA simulation results.
In {\bf BD2b} model, for example,
at $t_{\rm age}=33$~Myr (red dotted line in Fig. 5), these spectral indexes are
$A_{355}^{1413} \approx 0.85$ and $A_{1413}^{10000} \approx 1.07$.
In {\bf BD3b} model, again at $t_{\rm age}=33$~Myr, they are
$A_{355}^{1413} \approx 0.83$ and $A_{1413}^{10000} \approx 1.03$.
So if the shock age or the electron acceleration duration is about 30~Myr,
relatively young compared to the dynamical time scale of typical clusters, 
the curved radio spectrum around 1~GHz can be explained by DSA at cluster shocks.

In the case of {\bf BD3b model}, the spectral index behaves very similarly to that of a plane shock case
(see Fig. 4 of Paper 1), since the shock speed is more or less constant in time. 
Departures from the predictions for the test-particle planar shock are the most severe in {\bf BD1} ({\bf MF1}) model,
while it becomes relatively milder in {\bf BD2} and {\bf BD3} models with decreasing halo density.

As shown in Paper I, any variations in the spatial distributions of $f_e(r)$ and $B_d(r)$ are smoothed in
the volume integrated quantities such as $J_{\nu}$.
So signatures imprinted on the volume-integrated emission due to different $B_d(r)$ (e.g., between {\bf BD2} and 
{\bf BD2b} models) would be too subtle to detect.

\section{SUMMARY}

We have performed time-dependent DSA simulations for cosmic-ray (CR) electrons 
at decelerating spherical shocks with parameters relevant for weak cluster shocks:
$u_s\approx (3.0-4.5)\times 10^3 \kms$ and $M_s\approx 3.0-4.3$.
Several models with different postshock magnetic field profiles ({\bf MF1-3}) and  different upstream gas density
profiles ({\bf BD1-3}) were considered as summarized in Table 1.
Using the synchrotron emissivity, $j_{\nu}(r)$, calculated from the CR electron energy spectra at these model shocks, 
the radio surface brightness profile, $I_{\nu}(R)$, and the volume integrated spectrum, $J_{\nu}$,
were estimated by assuming a ribbon-like shock structure described in Fig. 1.

At low frequencies ($<1$~GHz) the surface brightness is affected by 
the evolution of $u_s(t)$ and the spatial variation of $B_d(r)$
as well as the projection effects ,
resulting in a wide range of spatial profiles of $I_{\nu}(R)$ (see Figs. 2 and 4).
At high frequencies ($>1$~GHz), such dependences become relatively weaker,
because the postshock width of high energy electrons is much narrower and so
the magnetic field profile far downstream has less influence on synchrotron emission.

For low frequency observations, the width of radio relics
increases with the shock age as $\Delta l_{\rm adv}\sim t_{\rm age} u_s/\sigma$, 
while it asymptotes to the cooling length,
$\Delta l_{\rm cool}\sim t_{\rm rad}(\gamma_e) u_s/\sigma$, for high frequencies.
If the surface brightness can be spatially resolved at multi-frequency observations
over $\sim (0.1-10)$~GHz, we may extract significant information about
the time evolution of $u_s(t)$, the shock age, $t_{\rm age}$,
and the postshock magnetic field strength, $B_d(r)$, 
through the detail modeling of DSA and projection effects.
On the contrary, the spectral index of $I_{\nu}(R)$ behaves rather similarly in all 
the models considered here. 

If the postshock magnetic field strength is about $7\muG$,
at the shock age of $\sim30$~Myr, the volume-integrated radio spectrum
has a break frequency, $\nu_{\rm br} \sim 1$~GHz, and 
steepens gradually with the spectral index from
$\alpha_{\rm inj}$ to $\alpha_{\rm inj}+0.5$ over the frequency range 
of 0.1-10~GHz (see Fig. 5).
Thus we suggest that such a curved spectrum could explain the observed
spectrum of the relic in cluster A2256 \citep{trasatti14}.

\acknowledgments{
This research was supported by Basic Science Research Program through the National Research Foundation of Korea(NRF) funded by the Ministry of Education (2014R1A1A2057940).
}


\end{document}